\documentclass[11pt,preprint,superscriptaddress,amsmath,amssymb,11pt,aps,showpacs,showkeys,nofootinbib,a4paper]{article}
\usepackage{amssymb,amsmath,amsfonts}
\usepackage{graphicx}
\usepackage{eepic,epsfig}
\usepackage{verbatim}
\usepackage{color}
\usepackage{hyperref}
\usepackage{cite}

1.60cm \makeatletter \@addtoreset{equation}{section}

\makeatother

\begin{document}
\title{Light-Cone Fluctuations in the Cosmic String Spacetime}
\author{H. F. Mota\thanks{E-mail: hmota@fisica.ufpb.br}, E. R. Bezerra de Mello\thanks
{E-mail: emello@fisica.ufpb.br}, C. H. G. Bessa\thanks{E-mail: carlos@cosmos.phy.tufts.edu} and V. B. Bezerra\thanks{E-mail: valdir@fisica.ufpb.br}\\
\\
\textit{Departamento de F\'{\i}sica, Universidade Federal da Para\'{\i}ba}\\
\textit{58.059-970, Caixa Postal 5.008, Jo\~{a}o Pessoa, PB, Brazil}\vspace{%
0.3cm}\\
}
\maketitle
%
\begin{abstract}
In this paper we consider light-cone fluctuations arising as a consequence of the nontrivial topology of the locally flat cosmic string spacetime. By setting the light-cone along the $z$-direction we are able to develop a full analysis to calculate the renormalized graviton two-point function, as well as the mean square fluctuation in the geodesic interval function and the time delay (or advance) in the propagation of a light-pulse. We found that all these expressions depend upon the parameter characterizing the conical topology of the cosmic string spacetime and vanish in the absence of it. We also point out that at large distances from the cosmic string the mean square fluctuation in the geodesic interval function is extremely small while in the opposite limit it logarithmically increases, improving the signal and thus, making possible the detection of such quantity.
\end{abstract}
\bigskip

\bigskip
%
\section{Introduction}
\label{Int}
%

Light-cone fluctuations have been an active topic of discussion in physics in the past few years and is one of the most relevant features that is expected to be exhibited in a complete theory of gravity. In fact, in a model of linearized quantum theory of gravity it has been shown that the effect of light-cone fluctuations is to smear out ultraviolet divergencies stemming from light-cone singularities of two-point functions \cite{Ford:1994cr}. This is in accordance with the conjecture made in 1956 by Pauli who said that active quantum fluctuations of spacetime metric might drive fluctuations of light-cones \cite{Pauli} which in turn could undertake the role of an universal regulator to remove quantum field theory divergencies (see \cite{Deser:1957zz, Isham:1970aw, Isham:1972pf} for further discussion). Moreover, light-cone fluctuations in the context of linearized quantum gravity model also offers a way of studying horizon fluctuations which may reveal new insights about black hole physics \cite{Ford:1997zb, Bekenstein:1995ju, Thompson:2008pqa, Thompson:2008vi}.

As it is common, by assuming that gravitons are in a squeezed vacuum state, the fluctuations in their propagation lead to a delay or advance in the time of propagation of a light-pulse toward its final destination, as a consequence of a nonzero linearized metric fluctuation responsible for inducing a nonzero averaged and finite Green's function taken on the light-cone. In this sense, the author in \cite{Ford:1994cr}, where the linearized quantum gravity model was developed, investigated gravitons in a flat spacetime and in an expanding universe (see also \cite{Ford:1996qc, Ford:1999xg}). Additionally, in Ref. \cite{ Yu:1999pq} fluctuations on the graviton's trajectory was investigated in flat spacetimes with nontrivial topology, and in Refs. \cite{Yu:1999wg, Yu:2000yf} the role of theories with extra dimensions was taken into account. The effect of compactified spacetimes on the light-cone fluctuations was also considered in Ref. \cite{Yu:2009mv}. In Refs. \cite{Hu:1997iu, Krein:2010ee, Arias:2011yg, Bessa:2012fs, DeLorenci:2012jq, Arias:2013mza} the authors studied metric and light-cone fluctuations using a stochastic approach.

Cosmic strings are linear topological defects arising due to phase transitions in the early universe and are predicted in the framework of some gauge extensions of the Standard Model of particle physics, possibly giving rise to a variety of cosmological, astrophysical and gravitational phenomena \cite{VS, hindmarsh, Copeland:2011dx}. From the gravitational point of view, for instance, the spacetime created by an indealized infinitely long and straight cosmic string presents a conical topology with a planar deficit angle given by $\Delta\phi = 8\pi G\mu$ on the plane perpendicular to it. Here $G$ is the Newton's gravitational constant and $\mu$ the cosmic string linear energy density.

The conical structure of the cosmic string spacetime disturbs the quantum vacuum fluctuations associated with scalar, fermionic and vector fields, providing that the vacuum expectation value of physical observables like the energy-momentum tensor \cite{PhysRevD.35.536, escidoc:153364, GL, DS, PhysRevD.46.1616, PhysRevD.35.3779, LB, Moreira1995365, BK, deMello:2011mw} or the Casimir-Polder force \cite{Saharian:2012pe, BezerradeMello:2012yf} is nonzero. By considering the presence of a magnetic flux running along the string, additional vacuum fluctuations associated with charged fields also take place  \cite{PhysRevD.36.3742, guim1994, SBM, SBM2, SBM3,Spinelly200477,SBM4,LS,SNDV,ERBM,PhysRevD.82.085033, Braganca:2014qma}.  Moreover, quantum gravity features have also been carried out in the context of the scattering of nonrelativistic and relativistic particles in $(2 + 1)$-dimensional cosmic string spacetime \cite{Deser:1988qn,Gerbert,Spinally:2000ii, Alvarez:1995fs, Mota:2016eoi}. In these works, the role of the cosmic string topology on the scattering amplitude was investigated. So, it is no surprise that the cosmic string nontrivial topology may also affect the fluctuations of the light-cone in such way there is a nonzero renormalized graviton two-point function. As we will see, the averaged graviton two-point function depends on the cosmic string parameter, $\alpha = 1 - 4G\mu$, and is responsible for producing a nonzero mean square fluctuation  (M.S.F) in the geodesic interval function which in turn yields a nonzero time delay (or advance) in the propagation of a light-pulse. Hence, the main objective of the present paper is to investigate the propagation of photons in the locally flat cosmic string spacetime in order to see how the (M.S.F) is affected by the cosmic string parameter $\alpha$.

The paper is organized as follows: In Sec. \ref{sec:lightcone} we review some general aspects of light-cone fluctuations following the approach suggested in Ref. \cite{Ford:1994cr}. In particular, we will see how the (M.S.F) depends on the renormalized graviton two-point function. In Sec. \ref{sec:string}, the latter is calculated in the cosmic string spacetime. In Sec. \ref{sec:lightcone-string}, we apply the two-point function found in Sec. \ref{sec:string} to obtain the (M.S.F) and derive the time delay (or advance) in the propagation of a light-pulse. Sec. \ref{sec:conc} is devoted to the conclusions and discussions. Some necessary calculations to obtain the results of Sec. \ref{sec:string} are presented in Appendices \ref{HF} and \ref{for}. Through the paper we work in natural units $\hbar = c = 1$.
\section{Light-cone fluctuations revisited}\label{sec:lightcone}
In this section we will review some aspects related to the light cone fluctuations approach. Let us then start by considering a line element in the form
\begin{equation}\label{metric1}
ds^2 = (\eta_{\mu\nu}^{(0)}+h_{\mu\nu})dx^\mu dx^\nu,
\end{equation}
where $\eta_{\mu\nu}^{(0)}$ is the metric tensor describing a flat spacetime, with $h_{\mu\nu}$ being its linearized perturbation. In the perturbed spacetime represented by the line element above, the half of the squared geodesic separation between two points $x$ and $x'$, defined as $\sigma(x, x')$, may be expanded in powers of $h_{\mu\nu}$, as it is shown below:
\begin{equation}\label{GS}
\sigma(x, x') \simeq \sigma_0(x, x') + \sigma_1(x, x'),
\end{equation}
where $2\sigma_0(x, x')=(x - x')^2 = (t -t')^2 - ({\bf x} - {\bf x}')^2$ is defined for the flat background and $\sigma_1(x, x')$ is only the first order term in the expansion.

On the other hand, by assuming that the first order perturbation metric tensor $h_{\mu\nu}(x)$ is quantized, its positive $h_{\mu\nu}^+(x)$ and negative $h_{\mu\nu}^-(x)$ frequencies decomposition will act on the squeezed vacuum state $|\psi\rangle$, such that $h_{\mu\nu}^+(x)|\psi\rangle = 0$ and $\langle\psi|h_{\mu\nu}^-(x) = 0$, straightforwardly providing $\langle\psi| h_{\mu\nu}|\psi\rangle = \langle h_{\mu\nu}\rangle = 0$. The metric fluctuations are, therefore, manifested through the calculation of the quantity $\langle h_{\mu\nu}^2\rangle$, which is in general nonzero.

In fact, a relation between $\langle h_{\mu\nu}^2\rangle$ and  $\langle\sigma_1^2\rangle$ follows from the null geodesic 
\begin{equation}\label{Nmetric}
dt^2 = d{\bf x}^2 - h_{\mu\nu}dx^\mu dx^\nu,
\end{equation}
which is obtained from Eq. (\ref{metric1}) using the transverse trace-free gauge, that is, $h_j^j=\partial_jh^{ij}=h^{0\nu}=0$. Thereby, the expansion of Eq. (\ref{Nmetric}) up to first order in $h_{\mu\nu}$ provides \cite{Ford:1994cr}
\begin{equation}\label{Nmetric2}
\Delta t = \Delta r - \frac{1}{2}\int_{r_0}^{r_1}h_{ij}n^in^jdr,
\end{equation}
where $dr=|d{\bf x}|$, $\Delta r=r_1-r_0$ and $n^i=dx^i/dr$ is a unit vector defining the spatial direction of the geodesic. 

Additionally, if one identifies the right hand side of Eq. (\ref{Nmetric2}) as being the proper spatial distance $\Delta\ell$ between two points in the spacetime, the square of the geodesic separation will be $2\sigma=\Delta t^2 - \Delta\ell^2$ and, as a consequence, expanding up to first order in $h_{\mu\nu}$, one obtains
\begin{equation}\label{Gsepa}
2\sigma \simeq \Delta t^2 - \Delta r^2 + \Delta r\int_{r_0}^{r_1}h_{ij}n^in^jdr.
\end{equation}
The correction to $\sigma_0$ is then give by the integral term in Eq. (\ref{Gsepa}), i.e
\begin{equation}\label{GS}
\sigma_1 = \frac{1}{2}\Delta r \int_{r_0}^{r_1}h_{\mu\nu}n^\mu n^\nu dr,
\end{equation}
which in turn also provides the vacuum expectation value
\begin{eqnarray}
\left\langle \sigma_1^2 \right\rangle_R =\frac{1}{8}\left(\Delta r\right)^2\int_{r_0}^{r_1}dr\int_{r_0}^{r_1}dr'n^in^jn^ln^m \left\langle h_{ij}(x)h_{lm}(x') + h_{ij}(x')h_{lm}(x) \right\rangle_R.
\label{eqsigma2}
\end{eqnarray}
The expression $\left\langle h_{ij}(x)h_{lm}(x') + h_{ij}(x')h_{lm}(x) \right\rangle_R$ is the renormalized graviton two-function and, as we can see, $h_{ij}(x)$ has a crucial role to calculate it. 

The light-cone fluctuations are codified in the propagation of a light-pulse which, because of boundary conditions or the topology of the spacetime, ends up to be delayed or advanced in time by an amount of $\Delta\tau$ given by
\begin{equation}\label{TdA}
\Delta\tau = \frac{\sqrt{\left\langle \sigma_1^2 \right\rangle_R}}{\Delta r}.
\end{equation}
Note that essentially, a nonzero value for $\left\langle \sigma_1^2 \right\rangle$ correspond to the fact that the retarded Green's function in flat spacetime for a massless scalar field has no longer a singularity at $\sigma_0=0$. This can be seen through the following expression for the averaged retarded Green's function \cite{Ford:1994cr} for a massless scalar field:
\begin{equation}\label{AGF}
\langle G_{\rm ret}(x,x')\rangle =\frac{\theta(t - t')}{8\pi^2}\sqrt{\frac{\pi}{2\left\langle \sigma_1^2 \right\rangle}}\exp\left(-\frac{\sigma_0^2}{2\left\langle \sigma_1^2 \right\rangle}\right),
\end{equation}
defined for $\left\langle \sigma_1^2 \right\rangle > 0$. It turns out that the result in Eq. (\ref{AGF}) is essential since the quantization of the metric perturbation leads, in the transverse trace-free gauge,  to a Klein-Gordon like equation, that is, $\Box h_{ij}=0$ \cite{Ford:1994cr, Ford:1977dj}. This means that the solution for $h_{ij}$ can be given in terms of a massless scalar field wave function having a plane wave-like solution. Nevertheless, when the line element describes a curved spacetime the Green's function is represented by the Hadamard function so that near the light-cone it has a flat leading asymptotic behaviour.

In the next section we will see how $h_{ij}$ can be evaluated in the cosmic string spacetime so that its influence in the fluctuations of the light-cone will be clear. We will also see that although the cosmic string spacetime is only locally flat, by setting the light-cone along the $z$-direction we will be able to confidently use Eq. (\ref{eqsigma2}) derived from Eq. (\ref{Nmetric}), which has a flat background metric.
%
\section{Graviton two-point function in the cosmic string spacetime}\label{sec:string}
\subsection{Massless scalar field in the cosmic string spacetime}
As it was said at the end of the previous section, an important point to quantify the metric perturbations in the transverse trace-free gauge is the massless scalar solution of the Klein-Gordon equation which will be obtained in this section. 

Let us then consider the line element describing the cosmic string spacetime, that is,
\begin{equation}
ds^2 = g_{\mu\nu}dx^{\mu}dx^{\nu}=dt^2 - d\rho^2 - \rho^2d\phi^2 - dz^2,
\label{LE}
\end{equation}
where the spacetime coordinates take values in the following interval: $\rho \geq 0$, $0 \leq \phi\leq \phi_0=2\pi/q$ and $-\infty\leq (t, z)\leq\infty$. Moreover, the parameter $q =1/\alpha$ is related to the presence of the cosmic string through $\alpha=1-4G\mu$, where $\mu$ is the linear energy density of the cosmic string and $G$ is the Newton's gravitational constant. Note that in order for the line element (\ref{LE}) to describe a cosmic string spacetime it is necessary to consider $q\geq 1$, otherwise, one would have a line element describing a disclination, i.e, in the case $0<q<1$ \cite{Katanaev:1992kh}. 

The field equation for a non-minimally coupled massless scalar field in a curved spacetime is given by the Klein-Gordon equation
\begin{eqnarray}
\left[\frac{1}{\sqrt{|g|}}\partial_{\rho}\left(\sqrt{|g|}g^{\rho\sigma}\partial_{\sigma}\right)+\xi \mathcal{R}\right]\Phi(x)=0,
\label{KGE}
\end{eqnarray}
being $g=\rm{det}(g_{\mu\nu})$, $\xi$ the non-minimal coupling constant to gravity and $\mathcal{R}$ the scalar curvature. In the cosmic string spacetime $\mathcal{R}=2(q-1)\delta(\rho)/\rho$. It is zero everywhere except at $\rho=0$, where the cosmic string is localized. However, as we aim to consider regions in space where $\rho>0$, the scalar curvature vanishes and, therefore, considering the line element (\ref{LE}), Eq. (\ref{KGE}) becomes
\begin{eqnarray}
\left[\frac{d^2}{d\rho^2} + \frac{1}{\rho}\frac{d}{d\rho}+\eta^2 - \frac{q^2n^2}{\rho^2}\right]f(\rho)=0,
\label{RKGE}
\end{eqnarray}
where we have used the ansatz,
\begin{eqnarray}
\Phi(x) = Ce^{-i\omega t + inq\phi + ik_zz}f(\rho),
\label{ansatz}
\end{eqnarray}
with $\eta^2=\omega^2 - k_z^2$, $C$ is a normalization constant and $f(\rho)$ an unknown radial function. As Eq. (\ref{RKGE}) is a Bessel differential equation, its regular solution at the origen is given by the Bessel function of the first kind, i.e, $f(\rho)=J_{q|n|}(\eta\rho)$. Thus, the general solution takes the form
\begin{eqnarray}
\Phi(x) = Ce^{-i\omega t + inq\phi + ik_zz}J_{q|n|}(\eta\rho).
\label{ansatz}
\end{eqnarray}

The constant $C$ can be obtained by the normalization condition 
\begin{eqnarray}
i\int d^3x\sqrt{|g|}\left[\Phi^{*}_{\gamma'}(x)\partial_t\Phi_{\gamma}(x) - \Phi_{\gamma}(x)\partial_t\Phi^{*}_{\gamma'}\right]=\delta_{\gamma,\gamma'},
\label{ansatz}
\end{eqnarray}
being $\gamma=(n,\eta,k_z)$ the set of quantum numbers and the delta symbol on the right-hand side is understood as Dirac delta function for the continuous quantum number, $\eta$ and $k_z$, and Kronecker delta for the discrete $n$. Thereby, we obtain 
\begin{eqnarray}
|C|^2 = \frac{q\eta}{8\pi^2\omega}.
\label{cons}
\end{eqnarray}
Therefore, the complete set of renormalized wave function is
\begin{eqnarray}
\Phi_{\gamma}(x) =\left( \frac{q\eta}{8\pi^2\omega}\right)^{\frac{1}{2}}e^{-i\omega t + inq\phi + ik_zz}J_{q|n|}(\eta\rho).
\label{RWF}
\end{eqnarray}

Having the solution above for the massless scalar field in the cosmic string spacetime we can proceed to calculate the graviton two-point function in the next section.
\subsection{Graviton two-point function}
As it has already been mentioned, the metric fluctuations can be written by means of a plane wave expansion of a massless scalar field. Thus, the general solution for $h_{ij}(x)$ is given by
\begin{equation}
h_{ij}(x) = \sum_{\gamma,\lambda}\left[a_{\gamma,\lambda}e_{ij}({\bf k},\lambda)\Phi_{\gamma}(x) + \rm{h.c.})\right],
\label{eq1}
\end{equation}
where ${\bf k} = (\eta, k_z)$ represents the wave vector in cylindrical coordinates, $\lambda$ labels the polarization states, $e_{ij}({\bf k},\lambda)$ is the polarization tensor and the sum over $\gamma$ means
\begin{equation}
\sum_{\gamma} = \int dk_z\int d\eta\sum_{n}.
\label{eq2}
\end{equation}
The massless scalar field  $\Phi_{\gamma}(x)$ satisfies the Klein-Gordon equation (\ref{KGE}) and, in the cosmic string spacetime, is given by Eq. (\ref{RWF}).

The graviton two-point function or, in other words, the Hadamard function is defined from Eq. (\ref{eqsigma2}) as
\begin{eqnarray}
G_{ijlm}(x,x') = \left\langle h_{ij}(x)h_{lm}(x') + h_{ij}(x')h_{lm}(x) \right\rangle,
\label{eq5}
\end{eqnarray}
which, by using the expression in Eq.  (\ref{eq1}) for $h_{ij}$, becomes 
\begin{eqnarray}
G_{ijlm}(x,x') = 2\mathrm{Re}\sum_{\gamma,\lambda}e_{ij}({\bf k},\lambda)e_{lm}({\bf k},\lambda)\Phi_{\gamma}(x)\Phi_{\gamma}^{*}(x').
\label{eq6}
\end{eqnarray}
Note that the two-point function above presents a singular behaviour at the coincidence limit, $x'\rightarrow x$, so that a renormalization procedure is needed to obtain a finite and well defined result. In this sense, a suitable renormalization procedure can be implemented by subtracting from $G_{ijlm}(x,x')$ the corresponding Minkowski contribution. 

An expression for the sum in $\lambda$ of the polarization tensors was obtained in \cite{ Yu:1999pq}, in cartesian coordinates, and is given by
\begin{eqnarray}
\sum_{\lambda}e_{ij}({\bf {k}},\lambda)e_{lm}({\bf {k}},\lambda) &=& \delta_{il}\delta_{jm} + \delta_{im}\delta_{jl} - \delta_{ij}\delta_{lm} + \hat{k}_i\hat{k}_j\hat{k}_l\hat{k}_m + \hat{k}_i\hat{k}_j\delta_{lm}\nonumber\\
&+& \hat{k}_l\hat{k}_m\delta_{ij} - \hat{k}_i\hat{k}_m\delta_{jl} - \hat{k}_i\hat{k}_l\delta_{jm} - \hat{k}_j\hat{k}_m\delta_{il} - \hat{k}_j\hat{k}_l\delta_{im},
\label{eq7}
\end{eqnarray}
with $\hat{k}_i=k_i/|{\bf k}|$ and $|{\bf k}| = \omega$. One should note that although the line element  (\ref{LE}) describing the cosmic string spacetime is given in cylindrical coordinates, the assumption of setting the light-cone along the $z$-direction allows us to adapt (\ref{eq7}) for our purpose. Thus, the only component of the unit vectors in Eq. (\ref{eqsigma2}) is $n^z$, so that the graviton Hadamard function (\ref{eq6}) simplifies to 
\begin{eqnarray}
G_{zzzz}(x,x') &=& 2\mathrm{Re}\sum_{\sigma}\left[1 - 2\frac{k_z^2}{|{\bf k}|^2} + \frac{k_z^4}{|{\bf k}|^4}\right]\Phi_{\gamma}(x)\Phi_{\gamma}^{*}(x'),\nonumber\\
&=&2\left(G^{(\rm{cs})}(x,x') - 2F_{zz}(x,x') + H_{zzzz}(x,x')\right),
\label{eq8}
\end{eqnarray}
being $G^{(\rm{cs})}(x,x')$ the propagator of a massless scalar field in the cosmic string spacetime and the functions $F_{zz}(x,x')$ and $H_{zzzz}(x,x') $ are defined as
\begin{eqnarray}
F_{zz}(x,x') &&=-\mathrm{Re}\sum_{\sigma}\frac{\partial_{\Delta z}^2}{|{\bf k}|^2}\Phi_{\gamma}(x)\Phi_{\gamma}^{*}(x'),
\label{eq11.1}
\end{eqnarray}
and
\begin{eqnarray}
H_{zzzz}(x,x') &=& \mathrm{Re}\sum_{\sigma}\frac{\partial_{\Delta z}^4}{|{\bf k}|^4}\Phi_{\gamma}(x)\Phi_{\gamma}^{*}(x'),
\label{eq12}
\end{eqnarray}
where $\partial_{\Delta z}\equiv\frac{\partial}{\partial\Delta z}$ and $\Delta z = z - z'$.

In the next section we will explicitly calculate the graviton Hadamard function (\ref{eq8}) with the help of Eqs. (\ref{eq11.1}) and (\ref{eq12}) which, together with  $G^{(\rm{cs})}(x,x')$, are also explicitly calculated in Appendices \ref{HF} and \ref{for}.
\section{Light-cone Fluctuation in the Cosmic String Spacetime}\label{sec:lightcone-string}
In this section we will consider the results presented in Appendices for the renormalized graviton Hadamard function, Eq. (\ref{B23}), obtained from Eq. (\ref{eq8}). These results will allow us to see the effects of the nontrivial topology of the cosmic string spacetime, described by the metric  (\ref{LE}), in the fluctuations of the light-cone. The latter manifests itself through a nonzero value for the expression in Eq. (\ref{TdA}), which represents a shift, an advance or delay, in the time of propagation of a light-pulse. Thus, let us consider the mean value of the square of the first order perturbation of the geodesic distance given by Eq. (\ref{eqsigma2}), i.e
\begin{eqnarray}
\left\langle \sigma_1^2 \right\rangle_R =\frac{1}{8}\left(b - a\right)^2\int_{a}^{b}dz\int_{a}^{b}dz'G^{(\rm{R})}_{zzzz}(\Delta t, \Delta z, \rho_0)|_{\Delta t = \Delta z},
\label{GS}
\end{eqnarray}
where we have considered the graviton wave propagation along the $z$-direction from $(t,\rho_0,\varphi_0,a)\rightarrow (t',\rho_0,\varphi_0,b)$ and $G^{(\rm{R})}_{zzzz}(\Delta t, \Delta z, \rho_0)|_{\Delta t = \Delta z}$ is given by Eq. (\ref{B23}) taken on the light-cone, with (\ref{B24}) written as
\begin{eqnarray}
G(\Delta t, \sigma, R, s)|_{\Delta t = \Delta z} &=& \frac{1}{6\pi^2R^8}\left(-3\Delta z^2s^4 + 94\Delta z^4s^2 - 8\Delta z^6\right)\nonumber\\
&-&\frac{\Delta z}{8\pi^2R^9}\ln\left(\frac{R + \Delta z}{R - \Delta z}\right)^2\left(-s^6 - 12\Delta z^2s^4 + 24\Delta z^4s^2\right).
\label{B24.1}
\end{eqnarray}
Note that $R$ and $s$ are given by (\ref{B10}) and (\ref{B26}), respectively. Note also that the above expression is similar to the one obtained in Ref. \cite{ Yu:1999pq}.

As the graviton Hadamard function (\ref{B23}) is an even function of $\Delta z$, by applying the Leibniz integral rule, Eq. (\ref{GS}) becomes \cite{arfken2011mathematical, Ford:1996qc}
\begin{eqnarray}
\left\langle \sigma_1^2 \right\rangle_R =\frac{1}{4}z_0^2\int_{0}^{z_0}dr(z_0 - r)G^{(\rm{R})}_{zzzz}(r, \rho_0),
\label{GS2}
\end{eqnarray}
being $G^{(\rm{R})}_{zzzz}(r, \rho_0) = G^{(\rm{R})}_{zzzz}(\Delta t, \Delta z, \rho_0)|_{\Delta t = \Delta z}$ and we have made the change $r = \Delta z$ and $z_0 = b - a$. Hence, by using Eq. (\ref{B23}) taken on the light-cone, the integral in Eq.  (\ref{GS2}) is found to be
\begin{eqnarray}
\left\langle \sigma_1^2 \right\rangle_R =\frac{z_0^2}{4}\left[ \sideset{}{'}\sum_{n=1}^{[q/2]} I_n(z_0, s_n) -\frac{q\sin(q\pi)}{2\pi}\int_{0}^{\infty}d\xi\frac{I_\xi(z_0, s_\xi)}{[\cosh(q\xi) - \cos(q\pi)]}\right], 
\label{GS3}
\end{eqnarray}
where $[q/2]$ represents the integer part of $q/2$, and the prime on the sign of summation means that in the case $q$ is an integer number the term $n=q/2$ should be taken with the coefficient $1/2$. We also have
\begin{eqnarray}
I(z_0, s) &=& \frac{(z_0^2 + s^2)^{\frac{1}{2}}(8z_0^4 + 25z_0^2s^2 + 14s^4) - (8z_0^5 + 8z_0^3s^2 + 3z_0s^4)\ln\left[\frac{(z_0^2 + s^2)^{\frac{1}{2}} - z_0}{s}\right]}{6\pi^2(z_0^2 + s^2)^{\frac{5}{2}}}\nonumber\\ 
&-& \frac{7}{3\pi^2},
\label{int}
\end{eqnarray}
with $s$ given by $s_n = 2\rho_0\sin(n\pi/q)$ for the first term on the right hand side of Eq. (\ref{GS3}) and by $s_\xi = 2\rho_0\cosh(\xi/2)$ for the second term, both expressions defined in Appendix \ref{for}. Thereby, Eq. (\ref{GS3}) is the most general closed expression for $\left\langle \sigma_1^2 \right\rangle_R $. The corresponding shift in time of a light-pulse propagating along the $z$-axis in the cosmic string spacetime is then written as 

\begin{equation}
\Delta\tau = \frac{\sqrt{\left\langle \sigma_1^2 \right\rangle_R}}{z_0}.
\label{ADT}
\end{equation}

We can additionally analyse $ \left\langle \sigma_1^2 \right\rangle_R$ in the limits $\rho_0\gg z_0$ and $\rho_0\ll z_0$. Thereby, let us first begin with the former case and consider Eq. (\ref{int}) in the form 
\begin{eqnarray}
I(x) &=& \frac{(x^2 + 1)^{\frac{1}{2}}(8x^4 + 25x^2 + 14) - (8x^5 + 8x^3 + 3x)\ln\left[(x^2 + 1)^{\frac{1}{2}} - x\right]}{6\pi^2(x^2 + 1)^{\frac{5}{2}}}\nonumber\\ 
&-& \frac{7}{3\pi^2}.
\label{int2}
\end{eqnarray}
Here we consider $I(x) = I(z_0,s)$ and $x=\frac{z_0}{s}$. By taking the limit $x\ll 1$, Eq. (\ref{int2}) reduces to
\begin{eqnarray}
I(x) \simeq\frac{32x^6}{45\pi^2} + O(x^7),
\label{appro}
\end{eqnarray}
which is a valid approximation for both $I(x)$'s in the sum and in the integral on the right hand side of Eq. (\ref{GS3}). Hence, one gets
\begin{eqnarray}
\left\langle \sigma_1^2 \right\rangle_R &\simeq&\frac{z_0^2}{360\pi^2}\left(\frac{z_0}{\rho_0}\right)^{6}\left[ \sideset{}{'}\sum_{n=1}^{[q/2]} \frac{1}{\sin^6(n\pi/q)}\right.\nonumber\\
&-&\left.\frac{q\sin(q\pi)}{2\pi}\int_{0}^{\infty}d\xi\frac{1}{[\cosh(q\xi) - \cos(q\pi)]}\frac{1}{\cosh^{6}(\xi/2)}\right].
\label{GS4}
\end{eqnarray}
For integer values of $q$, we find
\begin{eqnarray}
\left\langle \sigma_1^2 \right\rangle_R& \simeq&\frac{z_0^2}{720\pi^2}\left(\frac{z_0}{\rho_0}\right)^{6}\sum_{n=1}^{q-1} \frac{1}{\sin^6(n\pi/q)}\nonumber\\
&=&\frac{z_0^2}{720\pi^2}\left(\frac{z_0}{\rho_0}\right)^{6}\frac{1}{945}(q^2 - 1)(2q^4 + 23q^2 + 191).
\label{GS5}
\end{eqnarray}
It is worth mentioning that the resulting expression above, obtained for integer values of $q$, is an analytic function and, thus, by analytic continuation, it is valid for all values of $q$. The result (\ref{GS5}) shows that for regions far way from the string, that is, $z_0\ll\rho_0$, the values of $\left\langle \sigma_1^2 \right\rangle_R$ are too small to be detected, since it decreases with $(z_0/\rho_0)^{6}$.

On the other hand, in order to analyse (\ref{GS3}) in the regime where $\rho_0\ll z_0$ it is useful to write Eq. (\ref{int}) as 
\begin{eqnarray}
I(y) &=& \frac{(y^2 + 1)^{\frac{1}{2}}(8 + 25y^2 + 14y^4)- (8 + 8y^3 + 3y^4)\ln\left[\frac{(y^2 + 1)^{\frac{1}{2}} - 1}{y}\right]}{6\pi^2(y^2 + 1)^{\frac{5}{2}}}\nonumber\\ 
&-& \frac{7}{3\pi^2}.
\label{int3}
\end{eqnarray}
Here we also consider $I(y)=I(z_0,s)$ and $y=\frac{s}{z_0}$. Thus, taking the limit $y\ll 1$, we obtain 
\begin{eqnarray}
I(y) &\simeq&-\frac{2}{3\pi^2} (3 + 4\ln(y/2)) + O(y^2),\nonumber\\
&\simeq&-\frac{2}{3\pi^2} (3 + 4\ln(\rho_0/z_0) + 4\ln(s/2\rho_0)),\nonumber\\
&\simeq&\frac{8}{3\pi^2}\left|\ln(\rho_0/z_0)\right|,
\label{appro2}
\end{eqnarray}
which is the dominant term in the expansion. This approximation is certainly valid for the first term on the right hand side of Eq. (\ref{GS3}). Nevertheless, one needs to be careful when applying it for the second term because the factor $s\rightarrow s_{\xi}=2\rho_0\cosh(\xi/2)$ varies up to infinity and, as a consequence, there is no guarantee that $y\ll 1$. However, since the integral in (\ref{GS3}) is an exponentially decaying function of $y$, we can consider the following additional approximation 
\begin{eqnarray}
\int_{0}^{\infty}d\xi\frac{I_\xi(z_0, s_\xi)}{[\cosh(q\xi) - \cos(q\pi)]}\leq I_0(z_0, s_0)\int_{0}^{\infty}d\xi\frac{1}{[\cosh(q\xi) - \cos(q\pi)]}.
\label{app}
\end{eqnarray}
%
%
\begin{figure}[!htb]
\begin{center}
\includegraphics[width=0.43\textwidth]{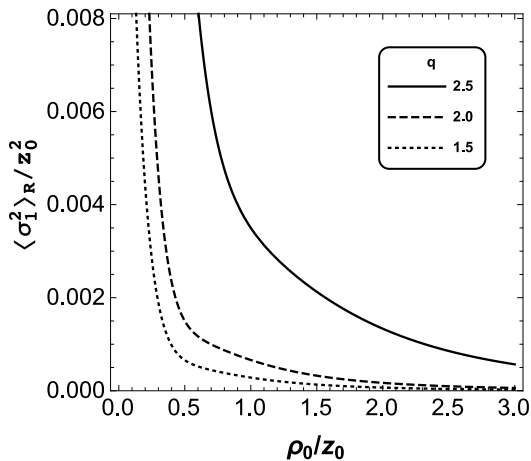}.
\caption{The square of the time shift, $\Delta\tau^2 =\left\langle \sigma_1^2 \right\rangle_R/z_0^2$, is plotted, as a function of $\rho_0/z_0$, for $q=1.5, 2.0$ and $2.5$.}
\label{fig1}
\end{center}
\end{figure}
This means that we can approximate $s_{\xi}$ by $2\rho_0$, providing that the approximation (\ref{appro2}) can also be adopted for the second term on the right hand side of Eq. (\ref{GS3}). One should note, however, that the approximation (\ref{app}) is only valid in the regime where $z_0\gg \rho_0$. Moreover, the error associated with this assumption is about $2\%$ for $\rho_0/z_0=0.001$ and $q=3/2$, and only decreases as $q$ increases or/and $\rho_0/z_0$ decreases. Therefore, we must write (\ref{GS3}) in the limit $\rho_0\ll z_0$ as
\begin{eqnarray}
\left\langle \sigma_1^2 \right\rangle_R &\simeq&\frac{2z_0^2}{3\pi^2}\left|\ln(\rho_0/z_0)\right|\left([q/2]'\right.\nonumber\\
&-&\left.\frac{q\sin(q\pi)}{2\pi}\int_{0}^{\infty}d\xi\frac{1}{[\cosh(q\xi) - \cos(q\pi)]}\right),
\label{GS6}
\end{eqnarray}
where the prime means that in the case $q$ is an integer $[q/2]\rightarrow (q-1)/2$. For the latter, Eq. (\ref{GS6}) becomes
\begin{eqnarray}
\left\langle \sigma_1^2 \right\rangle_R &\simeq&(q-1)\frac{z_0^2}{3\pi^2}\left|\ln(\rho_0/z_0)\right|,
\label{GS7}
\end{eqnarray}
which is also an analytical function of $q$, and by analytic continuation is valid for any value of $q$. The result above in the regime where $\rho_0\ll z_0$ is very interesting since it tell us that the values of $\left\langle \sigma_1^2 \right\rangle_R$ logarithmically increases as we consider points in the region near the cosmic string. Note that both expressions in Eqs. (\ref{GS5}) and (\ref{GS7}) vanish for $q=1$ as expected.

In Fig.\ref{fig1} we have plotted the general expression (\ref{GS3}) for the (M.S.F) as a function of $\rho_0/z_0$, in units of $z_0^2$, for $q=1.5, 2.0$ and 2.5. This quantity is also the square of the time shift, $\Delta\tau^2$. The plot reassure what we have already pointed out, i.e, considering $z_0$ fixed, for points far away from the cosmic string, the (M.S.F) decays with a power law of the form $(z_0/\rho_0)^{6}$, while for points near the cosmic string it logarithmically increases. It is interesting to note that, keeping $\rho_0$ fixed and increasing $z_0$, the time shift $\Delta\tau$ decreases, suggesting that over long flight distances the light-cone fluctuations tend to average to smaller values. We can also see that the values of $\left\langle \sigma_1^2 \right\rangle_R$ increase as $q$ is increased.
%
\section{Summary and Discussion}\label{sec:conc}
%
We have investigated the propagation of gravitons in the locally flat cosmic string spacetime by analyzing light-cone fluctuations arising due to the nontrivial topology. Following arguments of previous works \cite{Ford:1994cr, Ford:1977dj}, the general solution for the metric perturbation, $h_{ij}(x)$, in Eq. (\ref{eq1}), is given in terms of the solution of the massless scalar field. In this sense, we have then calculated the complete set of orthonormal solution (\ref{RWF}) of a massless scalar field by solving the Klein-Gordon equation in the cosmic string spacetime.

Because of the loss of isotropy of space, due to the presence of the cosmic string, we have considered the light-cone as being along the $z$-direction so that we have been able to obtain a general expression for the graviton two-point function. This expression is given in terms of the massless scalar field propagator $G^{(\rm cs)}(x,x')$ in the cosmic string spacetime and the functions $F_{zz}(x,x')$ and $H_{zzzz}(x,x')$, all of them found in Appendices and given by (\ref{A11}), (\ref{B11}) and (\ref{B22}), respectively. With these results we have calculated a closed expression for the renormalized graviton two-point function in Eq. (\ref{B23}), which in turn offered a way of obtaining a closed expression for the (M.S.F) found in Eqs. (\ref{GS3})-(\ref{int}), and consequently the delay or advance in time given by Eq. (\ref{ADT}), characterizing the light-cone fluctuations in the cosmic string spacetime.

Moreover, as the expression in (\ref{GS3}) is given in terms of an integral representation, two limiting cases were considered: when $\rho_0\gg z_0$ and when $z_0\gg \rho_0$. In the former limit we found the expression in Eq. (\ref{GS5}) for general values of $q$. The result in this case is neglectable since it is of order $(\rho_0/z_0)^6$. Regarding the case when $z_0\gg \rho_0$, using the additional reasonable approximation in Eq. (\ref{app}), we found the expression (\ref{GS7}), for general $q$. The result, in this regime, is much more interesting since (\ref{GS7}), or equivalently the time shift,  logarithmically increases with $\rho_0/z_0$. One should also note that all the results presented here are valid only for $\rho>0$, since at the origin, where the cosmic string is localized, there is a singularity. This behaviour can be clearly seen in Fig.\ref{fig1} which shows that (\ref{GS3}) logarithmically diverges as $\rho_0/z_0\rightarrow 0$.

Finally we would like to point out that, although there exist several works concerned with quantum field fluctuations in the cosmic string spacetime as mentioned in Sec. \ref{Int}, to the best of our knowledge, this is the first time an investigation about light-cone fluctuations in the cosmic string spacetime has been carried out.
%
\section*{Acknowledgments}
We are grateful to Larry Ford for valuable comments on the manuscript. We also would like to thank the Brazilian agency CNPq (Conselho Nacional de Desenvolvimento Cient\'{i}fico e Tecnol\'{o}gico - Brazil) for financial support. H.F.M is funded through the research project n$\textsuperscript{\underline{o}}$ 402056/2014-0. E.R.B.M is partially supported through the research project n$\textsuperscript{\underline{o}}$ 313137/2014-5. C.H.G.B is funded through the research project n$\textsuperscript{\underline{o}}$ 502029/2014-5. V.B.B is partially supported through the research project n$\textsuperscript{\underline{o}}$ 304553/2010-7.
\appendix
\section{Hadamard function in the cosmic string spacetime}
\label{HF}

The complete set of normalized mode functions given by Eq. (\ref{RWF}) allows us to evaluate the Hadamard function associated with the cosmic string spacetime as
\begin{equation}
G^{(\rm cs)}(x,x')=\sum_{\gamma}\Phi_{\sigma}(x)\Phi_{\gamma}^{*}(x'),
\label{A1}
\end{equation}
where $\gamma = (\eta,n,k_z)$ is the set of quantum numbers already introduced in Eq. (\ref{eq2}). Thereby, using (\ref{RWF}), Eq. (\ref{A1}) provides 
\begin{equation}
G^{(\rm cs)}(x,x')=\frac{q}{8\pi^2}\sum_{n=-\infty}^{\infty}\int_{-\infty}^{\infty}dk_ze^{ik_z\Delta z}\int_{0}^{\infty}d\eta\eta\frac{e^{i\omega\Delta t}}{\omega}J_{q|n|}(\eta\rho)J_{q|n|}(\eta\rho')e^{iqn\Delta\phi},
\label{A2}
\end{equation}
where $\Delta t=t-t'$, $\Delta z=z-z'$, $\Delta\varphi=\phi-\phi'$ and $\omega^2 = k_z^2 + \eta^2 + m^2$. Note that although we are interested in using the Hadamard function for the massless scalar field in the cosmic string spacetime we wish to go on calculating (\ref{A2}) as general as possible and only later on taking $m=0$.

The exponential term in the right-hand side of (\ref{A2}) can be written in the integral form
\begin{equation}
\frac{e^{\omega\Delta \tau}}{\omega}=\frac{2}{\sqrt{\pi}}\int_{0}^{\infty}dse^{-\omega^2s^2-\frac{\Delta\tau^2}{4s^2}},
\label{A3}
\end{equation}
where we have made a Wick rotation $\Delta\tau = i\Delta t$. The Hadamard function in the cosmic string spacetime now becomes
\begin{eqnarray}
G^{(\rm cs)}(x,x')&=&\frac{q}{4\pi^2\sqrt{\pi}}\int_{0}^{\infty}dse^{-m^2s^2-\frac{\Delta z^2}{4s^2}-\frac{\Delta\tau^2}{4s^2}}\int_{-\infty}^{\infty}dk_ze^{-s^2\left(k_z-\frac{i\Delta z}{2s^2}\right)^2}\nonumber\\
&&\times\sum_{n=-\infty}^{\infty}e^{iqn\Delta\phi}\int_{0}^{\infty}d\eta\eta e^{-\eta^2s^2}J_{q|n|}(\eta\rho)J_{q|n|}(\eta\rho').
\label{A4}
\end{eqnarray}
One can further simplify (\ref{A4}) using \cite{gradshteyn2000table2}
\begin{equation}
\int_{0}^{\infty}d\eta\eta e^{-\eta^2s^2}J_{q|n|}(\eta\rho)J_{q|n|}(\eta\rho') = \frac{e^{-\frac{(\rho^2 + \rho'^2)}{4s^2}}}{2s^2}I_{q|n|}\left(\frac{\rho\rho'}{2s^2}\right),
\label{A5}
\end{equation}
that is,
\begin{eqnarray}
G^{(\rm cs)}(x,x')=\frac{q}{8\pi^2\rho\rho'}\int_{0}^{\infty}dye^{-\frac{m\rho\rho'}{2y}-\frac{\Delta z^2y}{2\rho\rho'}-\frac{\Delta\tau^2y}{2\rho\rho'}-\frac{(\rho^2 + \rho'^2)y}{2\rho\rho'}}\times\sum_{n=-\infty}^{\infty}e^{iqn\Delta\phi}I_{q|n|}(y),
\label{A6}
\end{eqnarray}
where we have made the change $y=\rho\rho'/(2s^2)$. In order to solve the integral above we can make use of the summation formula derived previously in Ref. \cite{BezerradeMello:2011nv, deMello:2014ksa}, i.e
\begin{eqnarray}
\sum_{n=-\infty}^{\infty}e^{iqn\Delta\phi}I_{q|n|}(y) &=& \frac{e^y}{q} + \frac{2}{q}\sideset{}{'}\sum_{n=1}^{[q/2]}e^{y\cos\left(\frac{2\pi n}{q}-\Delta\phi\right)}\nonumber\\
&-& \frac{1}{2\pi}\sum_{j=+,-}\int_{0}^{\infty}d\xi\frac{\sin\left[q\left(j\Delta\phi + \pi\right)\right]e^{-y\cosh(\xi)}}{[\cosh(q\xi) - \cos(jq\Delta\phi + q\pi)]},
\label{A7}
\end{eqnarray}
where $[q/2]$ represents the integer part of $q/2$, and the prime on the sign of summation means that in the case $q$ is an integer number the term $n=q/2$ should be taken with the coefficient $1/2$. Note that, if $q<2$ the summation contribution should be omitted.

Hence, substituting (\ref{A7}) into (\ref{A6}) we obtain
\begin{eqnarray}
G^{(\rm cs)}(x,x')&=&\frac{m^2}{4\pi^2}\left[f_1(m\sigma_0) +2\sideset{}{'}\sum_{n=1}^{[q/2]} f_1(m\sigma_n)\right.\nonumber\\
&-&\left.\frac{q}{2\pi}\sum_{j=+,-}\int_{0}^{\infty}d\xi\frac{\sin\left[q\left(j\Delta\phi + \pi\right)\right]e^{-y\cosh(\xi)}}{[\cosh(q\xi) - \cos(jq\Delta\phi + q\pi)]} f_1(m\sigma_\xi)\right],
\label{A8}
\end{eqnarray}
where we have used the notation
\begin{eqnarray}
f_{\nu}(x) = \frac{K_{\nu}(x)}{x^{\nu}},
\label{A9}
\end{eqnarray}
with $K_{\nu}(x)$ being the modified Bessel function and
\begin{eqnarray}
\sigma_0^2 &=& -\Delta t^2 + \Delta z^2 + \rho^2 + \rho'^2 - 2\rho\rho'\cos(\Delta\phi),\nonumber\\
\sigma_n^2 &=& -\Delta t^2 + \Delta z^2 + \rho^2 + \rho'^2 - 2\rho\rho'\cos\left(\frac{2\pi n}{q}-\Delta\phi\right),\nonumber\\
\sigma_{\xi}^2 &=& -\Delta t^2 + \Delta z^2 + \rho^2 + \rho'^2 + 2\rho\rho'\cosh(\xi).
\label{A10}
\end{eqnarray}
Thus, Eq. (\ref{A8}) is a general closed expression for the Hadamard function in the cosmic string spacetime.

Taking now $m=0$ in Eq. (\ref{A6}), and using the summation formula (\ref{A7}) again we find
\begin{eqnarray}
G^{(\rm cs)}(x,x')&=&\frac{1}{4\pi^2}\frac{1}{\sigma_0^2} + \frac{1}{2\pi^2}\sideset{}{'}\sum_{n=1}^{[q/2]}\frac{1}{\sigma_n^2}\nonumber\\
&-&\frac{q}{8\pi^3}\sum_{j=+,-}\int_{0}^{\infty}d\xi\frac{\sin\left[q\left(j\Delta\phi + \pi\right)\right]e^{-y\cosh(\xi)}}{[\cosh(q\xi) - \cos(jq\Delta\phi + q\pi)]}\frac{1}{\sigma_\xi^2},
\label{A11}
\end{eqnarray}
which is the expression we use to calculate the graviton two-point function.

For integer values of $q$ the last term on the right-hand side of (\ref{A11}) vanishes and the summation in $n$ should be replaced with
\begin{eqnarray}
\sum_{n=1}^{[q/2]}\rightarrow\frac{1}{2}\sum_{n=1}^{q-1}.
\label{A12}
\end{eqnarray}
Thus, Eq. (\ref{A11})  reduces to 
\begin{eqnarray}
G^{(\rm cs)}(x,x')&=&\frac{1}{4\pi^2}\frac{1}{\sigma_0^2} + \frac{1}{4\pi^2}\sum_{n=1}^{q-1}\frac{1}{\sigma_n^2}.
\label{A13}
\end{eqnarray}

Therefore, Eqs. (\ref{A11}) and (\ref{A13}) are the expressions, when $q$ is general and integer, respectively, for the Hadamard function for a massless scalar field in the cosmic string spacetime. One should note that the renormalized propagators are obtained from Eqs.  (\ref{A8}) and (\ref{A11}) by subtracting the Minkowski contribution, which is the first term on the right hand side of each expression.
\section{Calculation of the functions $F_{zz}(x,x')$ and $H_{zzzz}(x,x')$}
\label{for}
In this appendix we wish to find a closed expression for the functions $F_{zz}(x,x')$ and $H_{zzzz}(x,x')$ given by (\ref{eq11.1}) and (\ref{eq12}), respectively. Let us then focus firstly on Eq. (\ref{eq11.1})  and write it in the form
\begin{eqnarray}
F_{zz}(x,x') &&=-\partial_{\Delta z}^2\mathrm{Re}\sum_{\gamma}\frac{e^{-i\omega\Delta t}}{\omega^3}\varphi_{\gamma}({\bf x})\varphi_{\gamma}^{*}({\bf x}'),
\label{eq11}
\end{eqnarray}
with $\varphi_{\gamma}({\bf x})$ being only the spatial part of the solution (\ref{RWF}), and we have taken $\omega$ out of the normalization constant. 

We would like now to proceed similarly to what we have done to calculate (\ref{A2}) by using the expression in Eq. (\ref{A3}). However, the eigenfrequency $\omega$ in the denominator of the above expression has a cubic power which makes the calculation more difficult. In order to overcome this problem, let us additionally consider the identity 
\begin{equation}
\frac{e^{-i\omega\Delta t}}{\omega^3} = -\int_0^{\Delta t}dt_2\int_0^{t_2}dt_1\frac{e^{-i\omega t_1}}{\omega} + \frac{1}{\omega^3} - \frac{i\Delta t}{\omega^2}.
\label{eq13}
\end{equation}
Thereby, upon substituting the identity (\ref{eq13}) into Eq. (\ref{eq11}), its real part is found to be
\begin{equation}
F_{zz}(x,x') =-\partial_{\Delta z}^2\left[-\int_0^{\Delta t}dt_2\int_0^{t_2}dt_1\sum_{\gamma}\frac{e^{-i\omega t_1}}{\omega}\varphi_{\gamma}({\bf x})\varphi_{\gamma}^{*}({\bf x}') + \sum_{\gamma}\frac{1}{\omega^3}\varphi_{\gamma}({\bf x})\varphi_{\gamma}^{*}({\bf x}')\right].
\label{eq15}
\end{equation}
Note that a similar `sum' over $\gamma$ in the first term on the right hand side of (\ref{eq15}) has already been developed in Appendix \ref{HF} and is given by (\ref{A11}), replacing $\Delta t$ with $t_1$. Regarding the second term on the right hand side, to carry it out, one can use the expression 
\begin{equation}
\frac{1}{\omega^{2s}}=\frac{2}{\Gamma(s)}\int_{0}^{\infty}d\tau \tau^{2s - 1}e^{-\omega^2\tau^2}.
\label{B4}
\end{equation}
Thus, following the same steps we took to get Eq.  (\ref{A6}), one has
\begin{equation}
\sum_{\sigma}\frac{\partial_{\Delta z}^2}{\omega^3}\varphi_{\gamma}({\bf x})\varphi_{\gamma}^{*}({\bf x}') = \frac{q}{8\pi^2}\partial_{\Delta z}^2\int_{0}^{\infty}\frac{dy}{y}e^{-\frac{\Delta z^2y}{2\rho\rho'}-\frac{(\rho^2 + \rho'^2)y}{2\rho\rho'}}\times\sum_{n=-\infty}^{\infty}e^{iqn\Delta\phi}I_{q|n|}(y).
\label{B5}
\end{equation}
Substituting the sum in $n$ given by (\ref{A7}), we can see that the integral in $y$ is logarithmically divergent at the origin. Nevertheless, we can introduce a positive regularization parameter, $p$, so that the integral can be solved as following:
\begin{eqnarray}
\int_{0}^{\infty}dy\frac{e^{-\frac{yR^2}{2\rho\rho'}}}{y} &=& \lim_{p\rightarrow 0} \int_{0}^{\infty}dy\frac{e^{-\frac{yR^2}{2\rho\rho'}}}{(y + p)}\nonumber\\
&=& \lim_{p\rightarrow 0}e^{\frac{R^2}{2\rho\rho'}p}\Gamma\left(0,\frac{R^2}{2\rho\rho'}p\right),
\label{B6}
\end{eqnarray}
where $\Gamma(a,z)$ is the incomplete gamma function. We can now expand, for small $p$, the right hand side of Eq. (\ref{B6}) as
\begin{eqnarray}
\partial^2_{\Delta z}\lim_{p\rightarrow 0}e^{xp}\Gamma\left(0,xp\right) &=& \partial^2_{\Delta z}\lim_{p\rightarrow 0}e^{xp}\left(-\gamma_e - \ln(xp) + p + O(p^2)\right)\nonumber\\
&=&-\partial^2_{\Delta z}\lim_{p\rightarrow 0}e^{xp}(\gamma_e + \ln(xp))\nonumber\\
&=&-\partial^2_{\Delta z}\lim_{p\rightarrow 0}\left(1+xp+O(p^2)\right)(\gamma_e + \ln(xp))\nonumber\\
&=&-\partial^2_{\Delta z}\lim_{p\rightarrow 0}\left[\gamma_e + \ln(x) + \ln(p) + \left(xp+O(p^2)\right) \ln(p)\right]\nonumber\\
&=&-\partial^2_{\Delta z}\ln(x),
\label{B7} 
\end{eqnarray}
being $x=R^2/(2\rho\rho')$ and $\gamma_e$ is the Euler's constant. Note that we have also exchanged the limit and the derivative so that $\partial_{\Delta z}(\gamma_e + \ln(p))=0$. In order to calculate Eq. (\ref{B5}) it is convenient to consider at this point that the wave is propagating along the $z$-direction from $(t,\rho_0,\varphi_0,z)\rightarrow (t',\rho_0,\varphi_0,z')$. Thus, with the result in (\ref{B7}), Eq. (\ref{B5}) becomes 
\begin{eqnarray}
\sum_{\sigma}\frac{\partial_{\Delta z}^2}{\omega^3}\varphi_{\gamma}({\bf x})\varphi_{\gamma}^{*}({\bf x}')&& =-\frac{1}{8\pi^2}\partial_{\Delta z}\left[\ln\left(\frac{\Delta z^2}{2\rho_0^2}\right) + 2\sideset{}{'}\sum_{n=1}^{[q/2]}\ln\left(\frac{R_n^2}{2\rho_0^2}\right) \right.\nonumber\\
&&\left.- \frac{q\sin(q\pi)}{\pi}\int_{0}^{\infty}d\xi\frac{\ln\left(\frac{R_\xi^2}{2\rho_0^2}\right)}{[\cosh(q\xi) - \cos(q\pi)]}\right],
\label{B9}
\end{eqnarray}
being the first term on the right hand side the Minkowski contribution and the others are the contributions due to the conical structure of the spacetime, with
\begin{eqnarray}
R_n^2 &=& \Delta z^2 + 4\rho_0^2\sin^2(\pi n/q),\nonumber\\
R_{\xi}^2 &=&  \Delta z^2 + 4\rho_0^2\cosh^2(\xi/2).
\label{B10}
\end{eqnarray}
Regarding the second term on the right hand side of (\ref{eq15}), as we have pointed out before, we can use Eq. (\ref{A11}) with $\Delta t\rightarrow t_1$. By integrating it we found 
\begin{eqnarray}
I_{zz} &=& \int_0^{\Delta t}dt_2\int_0^{t_2}dt_1\sum_{\gamma}\frac{e^{-i\omega t_1}}{\omega}\varphi_{\gamma}({\bf x})\varphi_{\gamma}^{*}({\bf x}'),\nonumber\\ 
&=& \frac{1}{4\pi^2}\left[S(\Delta t, \Delta z) - \frac{1}{2}\ln\left(\frac{\Delta z^2}{2\rho_0^2}\right) \right]\nonumber\\
&+&\frac{1}{2\pi^2}\sideset{}{'}\sum_{n=1}^{[q/2]}\left[S(\Delta t, R_n) - \ln\left(\frac{R_n^2}{2\rho_0^2}\right)\right]\nonumber\\
&-&\frac{q\sin(q\pi)}{4\pi^3}\int_{0}^{\infty}d\xi\frac{\left[S_\xi(\Delta t,R_\xi) - \ln\left(\frac{R_\xi^2}{2\rho_0^2}\right)\right]}{[\cosh(q\xi) - \cos(q\pi)]},
\label{B10.1}
\end{eqnarray}
where the first term on the right hand side is the Minkowski contribution and we use the general notation
\begin{equation}
S(\Delta t,R)=\left[\frac{\Delta t}{4R}\ln\left(\frac{R+\Delta t}{R-\Delta t}\right)^2+\frac{1}{2}\ln\left(\frac{R^2 -\Delta t^2}{2\rho_0^2}\right)\right].
\label{B12}
\end{equation}

Now, substituting the results (\ref{B9}) and (\ref{B10.1}) into Eq.  (\ref{B5}), we obtain 
\begin{equation}
F^{(\rm{R})}_{zz}(x,x') =\partial_{\Delta z}^2\left[\frac{1}{2\pi^2}\sideset{}{'}\sum_{n=1}^{[q/2]}S_n(\Delta t,R_n)-\frac{q\sin(q\pi)}{4\pi^3}\int_{0}^{\infty}d\xi\frac{S_\xi(\Delta t,R_\xi)}{[\cosh(q\xi) - \cos(q\pi)]}\right],
\label{B11}
\end{equation}
where we have subtracted the Minkowski contribution, which is the divergent contribution on the light-cone and needs to be removed. Note that, for integer values of $q$, the second term on the right hand side of (\ref{B11}) vanishes.

Let us now turn to the calculation of the function $H_{zzzz}(x,x')$. Thus, similarly to Eq. (\ref{eq11}), it can be written as
\begin{eqnarray}
H_{zzzz}(x,x') =\partial_{\Delta z}^4\mathrm{Re}\sum_{\gamma}\frac{e^{-i\omega\Delta t}}{\omega^5}\varphi_{\gamma}({\bf x})\varphi_{\gamma}^{*}({\bf x}').
\label{B13}
\end{eqnarray}
In order to evaluate (\ref{B13}) , we consider the following identity:
\begin{equation}
\frac{e^{-i\omega\Delta t}}{\omega^5} =\int_0^{\Delta t}dt_4\int_0^{t_4}dt_3 \int_0^{ t_3}dt_2\int_0^{t_2}dt_1\frac{e^{-i\omega t_1}}{\omega} + \frac{1}{\omega^5} - \frac{\Delta t^2}{2\omega^3} - \frac{i\Delta t}{2\omega^4} + \frac{i\Delta t^3}{6\omega^2}.
\label{B14}
\end{equation}
Substituting (\ref{B14}) into (\ref{B13}), its real part is given by
\begin{eqnarray}
H_{zzzz}(x,x') &=&\partial_{\Delta z}^4\left[\int_0^{\Delta t}dt_4\int_0^{t_4}dt_3 \int_0^{ t_3}dt_2\int_0^{t_2}dt_1\sum_{\gamma}\frac{e^{-i\omega t_1}}{\omega}\varphi_{\gamma}({\bf x})\varphi_{\gamma}^{*}({\bf x}')\right.\nonumber\\
&&\left.- \frac{\Delta t^2}{2}\sum_{\gamma}\frac{1}{\omega^3}\varphi_{\gamma}({\bf x})\varphi_{\gamma}^{*}({\bf x}') +  \sum_{\gamma}\frac{1}{\omega^5}\varphi_{\gamma}({\bf x})\varphi_{\gamma}^{*}({\bf x}')\right].
\label{B15}
\end{eqnarray}
Here again the `sum' over $\gamma$ in the first term on the right hand side is given by (\ref{A11}), with $\Delta t\rightarrow t_1$. Moreover, the second term on the right hand side has already been obtained and is given by Eq. (\ref{B9}) and the third term can be worked out similarly. Hence, by using (\ref{B4}), the latter can be written as
\begin{equation}
\sum_{\gamma}\frac{\partial_{\Delta z}^4}{\omega^5}\varphi_{\gamma}({\bf x})\varphi_{\gamma}^{*}({\bf x}') = \frac{q}{8\pi^2}\frac{\rho\rho'}{3}\partial_{\Delta z}^4\int_{0}^{\infty}\frac{dy}{y^2}e^{-\frac{\Delta z^2y}{2\rho\rho'}-\frac{(\rho^2 + \rho'^2)y}{2\rho\rho'}}\times\sum_{n=-\infty}^{\infty}e^{iqn\Delta\phi}I_{q|n|}(y).
\label{B16}
\end{equation}
The sum in $n$ is given by Eq.  (\ref{A7}) and the integral in $y$ above is again divergent. Nevertheless, as before, we can introduce a regularization parameter so that the divergent integral can be solved as
\begin{eqnarray}
\int_{0}^{\infty}dy\frac{e^{-\frac{yR^2}{2\rho\rho'}}}{y^2} &=& \lim_{p\rightarrow 0} \int_{0}^{\infty}dy\frac{e^{-\frac{yR^2}{2\rho\rho'}}}{(y + p)^2}\nonumber\\
&=& \lim_{p\rightarrow 0}\left[\frac{1}{p}-\frac{R^2}{2\rho\rho'}e^{\frac{R^2}{2\rho\rho'}p}\Gamma\left(0,\frac{R^2}{2\rho\rho'}p\right)\right].
\label{B17}
\end{eqnarray}
Following the same steps as before, the limit in Eq. (\ref{B17}) is found to be
\begin{eqnarray}
\partial^4_{\Delta z}\lim_{p\rightarrow 0}\left[\frac{1}{p}-xe^{xp}\Gamma\left(0,xp\right)\right] =\partial^4_{\Delta z}(x\ln(x)).
\label{B18} 
\end{eqnarray}
Once again taking the wave propagation in the $z$-direction so that $(t,\rho_0,\varphi_0,z)\rightarrow (t',\rho_0,\varphi_0,z')$, and using Eq. (\ref{B18}), the expression in (\ref{B16}) turns into
\begin{eqnarray}
\sum_{\gamma}\frac{\partial_{\Delta z}^4}{\omega^5}\varphi_{\gamma}({\bf x})\varphi_{\gamma}^{*}({\bf x}') &=& \frac{1}{48\pi^2}\partial_{\Delta z}^4\left[\Delta z^2\ln\left(\frac{\Delta z^2}{2\rho_0^2}\right) + 2\sideset{}{'}\sum_n^{[q/2]}R_n^2\ln\left(\frac{R_n^2}{2\rho_0^2}\right)\right.\nonumber\\
&&\left. - \frac{q\sin(q\pi)}{\pi}\int_{0}^{\infty}d\xi\frac{R_\xi^2\ln\left(\frac{R_\xi^2}{2\rho_0^2}\right)}{[\cosh(q\xi) - \cos(q\pi)]}\right],
\label{B19}
\end{eqnarray}
where the first term represents the Minkowski contribution.

On the other hand, using (\ref{A11}), the integral of the first term on the right hand side of Eq. (\ref{B15}) can be written 
\begin{eqnarray}
I_{zzzz}&=&\int_0^{\Delta t}dt_4\int_0^{t_4}dt_3 \int_0^{ t_3}dt_2\int_0^{t_2}dt_1\sum_{\gamma}\frac{e^{-i\omega t_1}}{\omega}\varphi_{\gamma}({\bf x})\varphi_{\gamma}^{*}({\bf x}')\nonumber\\
&=&\frac{1}{48\pi^2} \left[M(\Delta t, \Delta z) - (3\Delta t^2 + \Delta z^2)\ln\left(\frac{\Delta z^2}{2\rho_0^2}\right) - 5\Delta t^2\right]\nonumber\\
&+&\frac{1}{24\pi^2}\sideset{}{'}\sum_{n=1}^{[q/2]} \left[M_n(\Delta t,R_n) - (3\Delta t^2 + R_n^2)\ln\left(\frac{R_n^2}{2\rho_0^2}\right) - 5\Delta t^2\right]\nonumber\\
&-&\frac{q\sin(q\pi)}{48\pi^3}\int_{0}^{\infty}d\xi\frac{\left[M_\xi(\Delta t,R_\xi) - (3\Delta t^2 + R_\xi^2)\ln\left(\frac{R_\xi^2}{2\rho_0^2}\right)- 5\Delta t^2\right]}{[\cosh(q\xi) - \cos(q\pi)]},
\label{B20}
\end{eqnarray}
being the first term on the right hand side the Minkowski contribution and we use the general notation 
\begin{equation}
M(\Delta t,R)=\left(3R\Delta t + \frac{\Delta t^3}{R}\right)\ln\left(\frac{R + \Delta t}{R - \Delta t}\right) + (3\Delta t^2 + R^2)\ln\left(\frac{R^2 - \Delta t^2}{2\rho_0^2}\right). 
\label{B21}
\end{equation}

By substituting Eqs. (\ref{B9}),  (\ref{B19}) and (\ref{B20}) into Eq. (\ref{B15}), the renormalized expression is written as
\begin{eqnarray}
H^{(\rm{R})}_{zzzz}(x,x') &=&\partial_{\Delta z}^4\left[\frac{1}{24\pi^2}\sideset{}{'}\sum_{n=1}^{[q/2]} M_n(\Delta t,R_n)\right.\nonumber\\
&&\left. -\frac{q\sin(q\pi)}{48\pi^3}\int_{0}^{\infty}d\xi\frac{M_\xi(\Delta t,R_\xi)}{[\cosh(q\xi) - \cos(q\pi)]}
\right],
\label{B22}
\end{eqnarray}
where we have subtracted the Minkowski contribution. Note that because of the derivative in $\Delta z$ the terms with $5\Delta t^2$ in Eq. (\ref{B20}) have been neglected in Eq. (\ref{B22}). 

Once the functions $G^{(\rm cs)}(x,x')$, $F^{(\rm{R})}_{zz}(x,x')$ and $H^{(\rm{R})}_{zzzz}(x,x')$ have been calculated, after taking the derivatives with respect to $\Delta z$ in Eqs. (\ref{B11}) and (\ref{B22}), a closed expression for Eq. (\ref{eq8}) is found to be
\begin{eqnarray}
G^{(\rm{R})}_{zzzz}(\Delta t, \Delta z, \rho_0) =  \sideset{}{'}\sum_{n=1}^{[q/2]}G_n(\Delta t, \sigma_n, R_n, s_n)-\frac{q\sin(q\pi)}{2\pi}\int_{0}^{\infty}d\xi\frac{G_\xi(\Delta t, \sigma_\xi, R_\xi, s_\xi)}{[\cosh(q\xi) - \cos(q\pi)]},
\label{B23}
\end{eqnarray}
where 
\begin{eqnarray}
G(\Delta t, \sigma, R, s) &=& \frac{1}{6\pi^2R^8\sigma^2}\left[\left(\Delta z^2 - \Delta t^2\right)\left(16\Delta z^6 - 24\Delta z^4\Delta t^2\right) - 3\Delta t^2s^6\right.\nonumber\\
&+&\left.\left(9\Delta t^4 + 69\Delta z^2\Delta t^2 + 16\Delta z^4\right)s^4 +  \left(32\Delta z^6 + 32\Delta z^4\Delta t^2 - 72\Delta z^2\Delta t^4\right)s^2\right]\nonumber\\
&-&\frac{\Delta t}{8\pi^2R^9}\ln\left(\frac{R + \Delta t}{R - \Delta t}\right)^2\left[-s^6 - \left(3\Delta t^2 + 9\Delta z^2\right)s^4 + 24\Delta t^2\Delta z^2 s^2 \right.\nonumber\\
&&\left. - 8\Delta z^4\Delta t^2 + 8\Delta z^6\right],
\label{B24}
\end{eqnarray}
with
\begin{eqnarray}
\sigma_n &=& -\Delta t^2 + R_n^2,\nonumber\\
\sigma_\xi &=& -\Delta t^2 + R_\xi^2,
\label{B25}
\end{eqnarray}
and 
\begin{eqnarray}
s_n = 2\rho_0\sin(n\pi/q),\nonumber\\
s_\xi = 2\rho_0\cosh(\xi/2).
\label{B26}
\end{eqnarray}
The results derived in these appendices are applied to our analysis through the body of the text.
%
%


\begin{thebibliography}{10}

\bibitem{Ford:1994cr}
L.~H. Ford, {\it {Gravitons and light cone fluctuations}},  {\em Phys. Rev.}
  {\bf D51} (1995) 1692--1700,
  [\href{http://xxx.lanl.gov/abs/gr-qc/9410047}{{\tt gr-qc/9410047}}].

\bibitem{Pauli}
W.~Pauli, {\em Helv. Phys. Acta. Suppl.} {\bf 4} (1956), no.~69.

\bibitem{Deser:1957zz}
S.~Deser, {\it {General Relativity and the Divergence Problem in Quantum Field
  Theory}},  {\em Rev. Mod. Phys.} {\bf 29} (1957) 417.

\bibitem{Isham:1970aw}
C.~J. Isham, A.~Salam, and J.~A. Strathdee, {\it {Infinity suppression gravity
  modified quantum electrodynamics}},  {\em Phys. Rev.} {\bf D3} (1971)
  1805--1817.

\bibitem{Isham:1972pf}
C.~J. Isham, A.~Salam, and J.~A. Strathdee, {\it {Infinity suppression in
  gravity modified electrodynamics. II}},  {\em Phys. Rev.} {\bf D5} (1972)
  2548--2565.

\bibitem{Ford:1997zb}
L.~H. Ford and N.~F. Svaiter, {\it {Cosmological and black hole horizon
  fluctuations}},  {\em Phys. Rev.} {\bf D56} (1997) 2226--2235,
  [\href{http://xxx.lanl.gov/abs/gr-qc/9704050}{{\tt gr-qc/9704050}}].

\bibitem{Bekenstein:1995ju}
J.~D. Bekenstein and V.~F. Mukhanov, {\it {Spectroscopy of the quantum black
  hole}},  {\em Phys. Lett.} {\bf B360} (1995) 7--12,
  [\href{http://xxx.lanl.gov/abs/gr-qc/9505012}{{\tt gr-qc/9505012}}].

\bibitem{Thompson:2008pqa}
R.~T. Thompson and L.~H. Ford, {\it {Enhanced Geometry Fluctuations in
  Minkowski and Black Hole Spacetimes}},  {\em Class. Quant. Grav.} {\bf 25}
  (2008) 154006, [\href{http://xxx.lanl.gov/abs/0802.1546}{{\tt
  arXiv:0802.1546}}].

\bibitem{Thompson:2008vi}
R.~T. Thompson and L.~H. Ford, {\it {Enhanced Black Hole Horizon
  Fluctuations}},  {\em Phys. Rev.} {\bf D78} (2008) 024014,
  [\href{http://xxx.lanl.gov/abs/0803.1980}{{\tt arXiv:0803.1980}}].

\bibitem{Ford:1996qc}
L.~H. Ford and N.~F. Svaiter, {\it {Gravitons and light cone fluctuations. 2:
  Correlation functions}},  {\em Phys. Rev.} {\bf D54} (1996) 2640--2646,
  [\href{http://xxx.lanl.gov/abs/gr-qc/9604052}{{\tt gr-qc/9604052}}].

\bibitem{Ford:1999xg}
L.~H. Ford, {\it {Space-time metric and light cone fluctuations}},  {\em Int.
  J. Theor. Phys.} {\bf 38} (1999) 2941--2958.

\bibitem{Yu:1999pq}
H.-W. Yu and L.~H. Ford, {\it {Light cone fluctuations in flat space-times with
  nontrivial topology}},  {\em Phys. Rev.} {\bf D60} (1999) 084023,
  [\href{http://xxx.lanl.gov/abs/gr-qc/9904082}{{\tt gr-qc/9904082}}].

\bibitem{Yu:1999wg}
H.-W. Yu and L.~H. Ford, {\it {Light cone fluctuations in quantum gravity and
  extra dimensions}},  {\em Phys. Lett.} {\bf B496} (2000) 107--112,
  [\href{http://xxx.lanl.gov/abs/gr-qc/9907037}{{\tt gr-qc/9907037}}].

\bibitem{Yu:2000yf}
H.-W. Yu and L.~H. Ford, {\it {Quantum light cone fluctuations in theories with
  extra dimensions}},  \href{http://xxx.lanl.gov/abs/gr-qc/0004063}{{\tt
  gr-qc/0004063}}.

\bibitem{Yu:2009mv}
H.~W. Yu, N.~F. Svaiter, and L.~H. Ford, {\it {Quantum Lightcone Fluctuations
  in Compactified Spacetimes}},  {\em Phys. Rev.} {\bf D80} (2009) 124019,
  [\href{http://xxx.lanl.gov/abs/0904.1087}{{\tt arXiv:0904.1087}}].

\bibitem{Hu:1997iu}
B.~L. Hu and K.~Shiokawa, {\it {Wave propagation in stochastic space-times:
  Localization, amplification and particle creation}},  {\em Phys. Rev.} {\bf
  D57} (1998) 3474--3483, [\href{http://xxx.lanl.gov/abs/gr-qc/9708023}{{\tt
  gr-qc/9708023}}].

\bibitem{Krein:2010ee}
G.~Krein, G.~Menezes, and N.~F. Svaiter, {\it {Analog model for quantum gravity
  effects: Phonons in random fluids}},  {\em Phys. Rev. Lett.} {\bf 105} (2010)
  131301, [\href{http://xxx.lanl.gov/abs/1006.3350}{{\tt arXiv:1006.3350}}].

\bibitem{Arias:2011yg}
E.~Arias, G.~Krein, G.~Menezes, and N.~F. Svaiter, {\it {Thermal Radiation from
  a Fluctuating Event Horizon}},  {\em Int. J. Mod. Phys.} {\bf A27} (2012)
  1250129, [\href{http://xxx.lanl.gov/abs/1109.6080}{{\tt arXiv:1109.6080}}].

\bibitem{Bessa:2012fs}
C.~H.~G. Bessa, J.~G. Duenas, and N.~F. Svaiter, {\it {Accelerated detectors in
  Dirac vacuum: the effects of horizon fluctuations}},  {\em Class. Quant.
  Grav.} {\bf 29} (2012) 215011, [\href{http://xxx.lanl.gov/abs/1204.0022}{{\tt
  arXiv:1204.0022}}].

\bibitem{DeLorenci:2012jq}
V.~A. De~Lorenci, G.~Menezes, and N.~F. Svaiter, {\it {Light-cone fluctuations
  and the renormalized stress tensor of a massless scalar field}},  {\em Int.
  J. Mod. Phys.} {\bf A28} (2013) 1350001,
  [\href{http://xxx.lanl.gov/abs/1208.3860}{{\tt arXiv:1208.3860}}].

\bibitem{Arias:2013mza}
E.~Arias, C.~H.~G. Bessa, J.~G. Dueñas, G.~Menezes, and N.~F. Svaiter, {\it
  {Casimir Energy Corrections by Light-Cone Fluctuations}},  {\em Int. J. Mod.
  Phys.} {\bf A29} (2014), no.~5 1450024,
  [\href{http://xxx.lanl.gov/abs/1307.4749}{{\tt arXiv:1307.4749}}].

\bibitem{VS}
A.~Vilenkin and E.~P.~S. Shellard, {\em {Cosmic strings and other topological
  defects}}.
\newblock Cambridge monographs on mathematical physics. Cambridge Univ. Press,
  Cambridge, 1994.

\bibitem{hindmarsh}
M.~Hindmarsh and T.~Kibble, {\it {Cosmic strings}},  {\em Rept.Prog.Phys.} {\bf
  58} (1995) 477--562, [\href{http://xxx.lanl.gov/abs/hep-ph/9411342}{{\tt
  hep-ph/9411342}}].

\bibitem{Copeland:2011dx}
E.~J. Copeland, L.~Pogosian, and T.~Vachaspati, {\it {Seeking String Theory in
  the Cosmos}},  {\em Class.Quant.Grav.} {\bf 28} (2011) 204009,
  [\href{http://xxx.lanl.gov/abs/1105.0207}{{\tt arXiv:1105.0207}}].

\bibitem{PhysRevD.35.536}
B.~Linet, {\it {Quantum Field Theory in the Space-time of a Cosmic String}},
  {\em Phys.Rev.} {\bf D35} (1987) 536--539.

\bibitem{escidoc:153364}
B.~Allen and E.~P.~S. Shellard, {\it {On the evolution of cosmic strings}},  in
  {\em {The formation and evolution of cosmic strings : proceedings of a
  workshop supported by the SERC and held in Cambridge, 3-7 July, 1989}} (G.~W.
  Gibbons, S.~W. Hawking, and T.~Vachaspati, eds.), (Cambridge), pp.~421--448,
  Cambridge University Press, 1990.

\bibitem{GL}
M.~Guimaraes and B.~Linet, {\it {Selfinteraction and quantum effects near a
  point mass in three-dimensional gravitation}},  {\em Class.Quant.Grav.} {\bf
  10} (1993) 1665--1680.

\bibitem{DS}
P.~Davies and V.~Sahni, {\it {Quantum gravitational effects near cosmic
  strings}},  {\em Class.Quant.Grav.} {\bf 5} (1988) 1.

\bibitem{PhysRevD.46.1616}
T.~Souradeep and V.~Sahni, {\it {Quantum effects near a point mass in
  (2+1)-Dimensional gravity}},  {\em Phys.Rev.} {\bf D46} (1992) 1616--1633,
  [\href{http://xxx.lanl.gov/abs/hep-ph/9208219}{{\tt hep-ph/9208219}}].

\bibitem{PhysRevD.35.3779}
V.~P. Frolov and E.~Serebryanyi, {\it {Vacuum Polarization in the Gravitational
  Field of a Cosmic String}},  {\em Phys.Rev.} {\bf D35} (1987) 3779--3782.

\bibitem{LB}
B.~Linet, {\it {Euclidean spinor Green's functions in the space-time of a
  straight cosmic string}},  {\em J.Math.Phys.} {\bf 36} (1995) 3694--3703,
  [\href{http://xxx.lanl.gov/abs/gr-qc/9412050}{{\tt gr-qc/9412050}}].

\bibitem{Moreira1995365}
J.~Moreira, E.S., {\it {Massive quantum fields in a conical background}},  {\em
  Nucl.Phys.} {\bf B451} (1995) 365--378,
  [\href{http://xxx.lanl.gov/abs/hep-th/9502016}{{\tt hep-th/9502016}}].

\bibitem{BK}
V.~B. Bezerra and N.~R. Khusnutdinov, {\it {Vacuum expectation value of the
  spinor massive field in the cosmic string space-time}},  {\em
  Class.Quant.Grav.} {\bf 23} (2006) 3449--3462,
  [\href{http://xxx.lanl.gov/abs/hep-th/0602048}{{\tt hep-th/0602048}}].

\bibitem{deMello:2011mw}
E.~R.~B. de~Mello, A.~A. Saharian, and A.~K. Grigoryan, {\it {Casimir Effect
  for Parallel Metallic Plates in Cosmic String Spacetime}},  {\em J. Phys.}
  {\bf A45} (2012) 374011, [\href{http://xxx.lanl.gov/abs/1111.7233}{{\tt
  arXiv:1111.7233}}].

\bibitem{Saharian:2012pe}
A.~A. Saharian and A.~S. Kotanjyan, {\it {Casimir?Polder potential for a
  metallic cylinder in cosmic string spacetime}},  {\em Phys. Lett.} {\bf B713}
  (2012) 133--139, [\href{http://xxx.lanl.gov/abs/1201.0135}{{\tt
  arXiv:1201.0135}}].

\bibitem{BezerradeMello:2012yf}
E.~R. Bezerra~de Mello, V.~B. Bezerra, H.~F. Mota, and A.~A. Saharian, {\it
  {Casimir-Polder interaction between an atom and a conducting wall in cosmic
  string spacetime}},  {\em Phys. Rev.} {\bf D86} (2012) 065023,
  [\href{http://xxx.lanl.gov/abs/1205.6914}{{\tt arXiv:1205.6914}}].

\bibitem{PhysRevD.36.3742}
J.~Dowker, {\it {Vacuum Averages for Arbitrary Spin Around a Cosmic String}},
  {\em Phys.Rev.} {\bf D36} (1987) 3742.

\bibitem{guim1994}
M.~Guimaraes and B.~Linet, {\it {Scalar Green's functions in an Euclidean space
  with a conical-type line singularity}},  {\em Commun.Math.Phys.} {\bf 165}
  (1994) 297--310.

\bibitem{SBM}
J.~Spinelly and E.~Bezerra~de Mello, {\it {Vacuum polarization of a charged
  massless scalar field on cosmic string space-time in the presence of a
  magnetic field}},  {\em Class.Quant.Grav.} {\bf 20} (2003) 873--888,
  [\href{http://xxx.lanl.gov/abs/hep-th/0301169}{{\tt hep-th/0301169}}].

\bibitem{SBM2}
J.~Spinelly and E.~Bezerra~de Mello, {\it {Vacuum polarization by a magnetic
  field in the cosmic string space-time}},  {\em Int.J.Mod.Phys.} {\bf A17}
  (2002) 4375--4384.

\bibitem{SBM3}
J.~Spinelly and E.~Bezerra~de Mello, {\it {Vacuum polarization of a charged
  massless fermionic field by a magnetic flux in the cosmic string
  space-time}},  {\em Int.J.Mod.Phys.} {\bf D13} (2004) 607--624,
  [\href{http://xxx.lanl.gov/abs/hep-th/0306103}{{\tt hep-th/0306103}}].

\bibitem{Spinelly200477}
J.~Spinelly and E.~Bezerra~de Mello, {\it {Vacuum polarization by a magnetic
  flux in a cosmic string background}},  {\em Nucl.Phys.Proc.Suppl.} {\bf 127}
  (2004) 77--83, [\href{http://xxx.lanl.gov/abs/hep-th/0305142}{{\tt
  hep-th/0305142}}].

\bibitem{SBM4}
J.~Spinelly and E.~Bezerra~de Mello, {\it {Spinor Green function in
  higher-dimensional cosmic string space-time in the presence of magnetic
  flux}},  {\em JHEP} {\bf 0809} (2008) 005,
  [\href{http://xxx.lanl.gov/abs/0802.4401}{{\tt arXiv:0802.4401}}].

\bibitem{LS}
L.~Sriramkumar, {\it {Fluctuations in the current and energy densities around a
  magnetic flux carrying cosmic string}},  {\em Class.Quant.Grav.} {\bf 18}
  (2001) 1015--1025, [\href{http://xxx.lanl.gov/abs/gr-qc/0011074}{{\tt
  gr-qc/0011074}}].

\bibitem{SNDV}
Y.~Sitenko and N.~Vlasii, {\it {Induced vacuum current and magnetic field in
  the background of a cosmic string}},  {\em Class.Quant.Grav.} {\bf 26} (2009)
  195009, [\href{http://xxx.lanl.gov/abs/0909.0405}{{\tt arXiv:0909.0405}}].

\bibitem{ERBM}
E.~R. Bezerra~de Mello, {\it {Induced fermionic current densities by magnetic
  flux in higher dimensional cosmic string spacetime}},  {\em
  Class.Quant.Grav.} {\bf 27} (2010) 095017,
  [\href{http://xxx.lanl.gov/abs/0907.4139}{{\tt arXiv:0907.4139}}].

\bibitem{PhysRevD.82.085033}
E.~Bezerra~de Mello, V.~Bezerra, A.~Saharian, and V.~Bardeghyan, {\it
  {Fermionic current densities induced by magnetic flux in a conical space with
  a circular boundary}},  {\em Phys.Rev.} {\bf D82} (2010) 085033,
  [\href{http://xxx.lanl.gov/abs/1008.1743}{{\tt arXiv:1008.1743}}].

\bibitem{Braganca:2014qma}
E.~A.~F. Bragança, H.~F. Santana~Mota, and E.~R. Bezerra~de Mello, {\it
  {Induced vacuum bosonic current by magnetic flux in a higher dimensional
  compactified cosmic string spacetime}},  {\em Int. J. Mod. Phys.} {\bf D24}
  (2015), no.~07 1550055, [\href{http://xxx.lanl.gov/abs/1410.1511}{{\tt
  arXiv:1410.1511}}].

\bibitem{Deser:1988qn}
S.~Deser and R.~Jackiw, {\it {Classical and Quantum Scattering on a Cone}},
  {\em Commun.Math.Phys.} {\bf 118} (1988) 495.

\bibitem{Gerbert}
P.~de~Sousa~Gerbert and R.~Jackiw, {\it {Classical and Quantum Scattering on a
  Spinning Cone}},  {\em Commun.Math.Phys.} {\bf 124} (1989) 229.

\bibitem{Spinally:2000ii}
J.~Spinally, E.~Bezerra~de Mello, and V.~Bezerra, {\it {Relativistic quantum
  scattering on a cone}},  \href{http://xxx.lanl.gov/abs/gr-qc/0012103}{{\tt
  gr-qc/0012103}}.

\bibitem{Alvarez:1995fs}
M.~Alvarez, F.~de~Carvalho~Filho, and L.~Griguolo, {\it {Time dependent quantum
  scattering in (2+1)-dimensional gravity}},  {\em Commun.Math.Phys.} {\bf 178}
  (1996) 467--482, [\href{http://xxx.lanl.gov/abs/hep-th/9507134}{{\tt
  hep-th/9507134}}].

\bibitem{Mota:2016eoi}
H.~Mota, {\it {Topological quantum scattering under the influence of a
  nontrivial boundary condition}},  {\em Mod. Phys. Lett.} {\bf A31} (2016),
  no.~11 1650074.

\bibitem{Ford:1977dj}
L.~H. Ford and L.~Parker, {\it {Quantized Gravitational Wave Perturbations in
  Robertson-Walker Universes}},  {\em Phys. Rev.} {\bf D16} (1977) 1601--1608.

\bibitem{Katanaev:1992kh}
M.~O. Katanaev and I.~V. Volovich, {\it {Theory of defects in solids and
  three-dimensional gravity}},  {\em Annals Phys.} {\bf 216} (1992) 1--28.

\bibitem{arfken2011mathematical}
G.~B. Arfken, H.~J. Weber, and F.~E. Harris, {\em Mathematical methods for
  physicists}.
\newblock Academic press, seventh edition~ed., 2013.

\bibitem{gradshteyn2000table2}
I.~Gradshteyn and I.~Ryzhik, {\it Table of integrals, series and products
  (corrected and enlarged edition prepared by d. zwillinger)},  {\em Academic
  Press} (2014).

\bibitem{BezerradeMello:2011nv}
E.~R. Bezerra~de Mello and A.~A. Saharian, {\it {Topological Casimir effect in
  compactified cosmic string spacetime}},  {\em Class. Quant. Grav.} {\bf 29}
  (2012) 035006, [\href{http://xxx.lanl.gov/abs/1107.2557}{{\tt
  arXiv:1107.2557}}].

\bibitem{deMello:2014ksa}
E.~R. Bezerra~de Mello, V.~B. Bezerra, A.~A. Saharian, and H.~H. Harutyunyan,
  {\it {Vacuum currents induced by a magnetic flux around a cosmic string with
  finite core}},  {\em Phys. Rev.} {\bf D91} (2015), no.~6 064034,
  [\href{http://xxx.lanl.gov/abs/1411.1258}{{\tt arXiv:1411.1258}}].

\end{thebibliography}

\end{document}